\input psfig.tex
\documentstyle[12pt,aasms4]{article}

\lefthead{Kemper, Spaans, Jansen, Hogerheijde, van Dishoeck, \& Tielens}
\righthead{Physical Models of Ced 201}

\begin{document}

\title{Far-Infrared and Sub-Millimeter Observations and Physical Models of the Reflection Nebula Ced 201}

\author{Ciska Kemper\footnote{Currently at the University of Amsterdam, Kruislaan 403, Amsterdam, The Netherlands}}

\affil{Leiden Observatory, P.O.~Box 9513, 2300 RA Leiden, The Netherlands}

\author{Marco Spaans\footnote{Hubble Fellow}}

\affil{Harvard-Smithsonian Center for Astrophysics, Cambridge, MA 02138, USA}

\author{David J.\ Jansen}

\affil{Leiden Observatory, P.O.~Box 9513, 2300 RA Leiden, The Netherlands}

\author{Michiel R.\ Hogerheijde\footnote{Currently at the University of
California in Berkeley, 601 Campbell Hall, CA 94720, USA}}

\affil{Leiden Observatory, P.O.~Box 9513, 2300 RA Leiden, The Netherlands}

\author{Ewine F.\ van Dishoeck}

\affil{Leiden Observatory, P.O.~Box 9513, 2300 RA Leiden, The Netherlands}

\author{Xander G.\ G.\ M.\ Tielens}

\affil{Kapteyn Institute, P.O.~Box 800, 9700 AV Groningen, The Netherlands}

\begin{abstract}
ISO [C II] 158 $\mu$m, [O I] 63 $\mu$m, and H$_2$ 9 and 17 $\mu$m
observations are presented of the reflection nebula
Ced 201, which is a photon-dominated region illuminated by a B9.5 star with
a color temperature of 10,000 K (a cool PDR).
In combination with ground based [C I] 609 $\mu$m,
CO, $^{13}$CO, CS and HCO$^+$
data, the carbon budget and physical structure of the reflection nebula are
constrained. The obtained data set is the first one to contain all important
cooling lines of a cool PDR, and allows a comparison to be made with classical
PDRs. To this effect one- and three-dimensional PDR models are
presented which incorporate the physical characteristics of the source, and are
aimed at understanding the dominant heating processes of the cloud.
The contribution of very small grains to the photo-electric heating rate is 
estimated from these models and used to constrain the total abundance of PAHs 
and small grains. Observations of the pure rotational H$_2$ lines with ISO, in
particular the S(3) line, indicate the presence of a small amount of very warm
$\sim 330$ K molecular gas. This gas cannot be accommodated by the presented
models.
\end{abstract}

{\it subject headings}: ISM: clouds - ISM: Ced 201 - ISM: molecules -
molecular processes - reflection nebulae

\section{Introduction}

A major problem concerning the interstellar medium and circumstellar 
regions remains the identification of the dominant gas heating source.
The best  current model is photo-electric emission by dust grains following
irradiation by ultraviolet photons. The ejected electrons carry an excess
kinetic energy of a few eV which is transferred to the gas by elastic
collisions with H and H$_2$.
The efficiency of this photo-electric heating for various 
densities and illuminating radiation fields has been the subject of many 
investigations (Hollenbach \& Tielens 1997; Spaans et al.~1994; and 
references therein). Bakes \& Tielens (1994) presented calculations 
for the contribution of Very Small Grains (VSGs) and large molecules like 
polycyclic hydrocarbons (PAHs) to the total heating rate. To establish a firm 
theoretical basis for interstellar heating, one therefore has to constrain the 
abundance of these putative species.

In this work, Infrared Space Observatory (ISO)
and ground-based observations are presented of the 
reflection nebula Ced 201. Since the illuminating star has an effective 
temperature of 10,000 K, Ced 201 qualifies as a cool PDR.
That is, the relatively soft impinging radiation field decreases the
photo-electric heating rate, leading to cooler temperatures. Furthermore,
the resulting photo-electric heating rate has larger contributions from
grain species with low ionization potentials and negative (or small
positive) charges (Bakes \& Tielens 1994). Ced 201 is thus
ideally suited to investigate the role of PAHs in the photo-electric
heating process (Spaans et al.~1994).

Most PDR models to date are 
constructed for classical PDRs, i.e. PDRs illuminated by O or early-type B 
stars. The aim of this study is to use a complete set of atomic and molecular 
diagnostics to constrain the total cooling, the density, the extinction, and 
the illuminating radiation field of a cool PDR. If the physical and chemical 
structure of the source can be determined consistent with the observations
through PDR models,
then it is possible to estimate the contribution of VSGs and PAHs to the total 
photo-electric heating rate.
ISO offers a new opportunity to probe the physics and chemistry of PDRs by
observations of the major cooling lines [C II] 158 $\mu$m and [O I] 63 $\mu$m
with the Long Wavelength Spectrometer (LWS) and searches for the pure
rotational lines of H$_2$ with the Short Wavelength Spectrometer (SWS). The
latter lines provide important constraints on the amount of warm gas in the
source, which may not be explaimed by current models (Timmerman et al.\
1996). Searches of these LWS and SWS lines in weak, extended sources such as
discussed here have not been possible previously with airborne observatories.

\section{Observations}

Cederblad 201 is associated with BD +69$^{\rm o}$1231, a main sequence star of 
spectral type B9.5 and an effective temperature of 10,000 K. The source is 
located toward $\alpha =$ 22:12:14 and $\delta =$ 70:00:11 (Epoch 1950) and 
lies at a distance of approximately 420 pc (Casey 1991). 
In this study an extended set of observations of Ced 201 is presented. The 
source is mapped around the offset position in $^{12}$CO 2 $\rightarrow$ 1 and 
3 $\rightarrow$ 2 line emission. Additional observations of line transitions 
of other species are performed at the maximum of the $^{12}$CO 2 $\rightarrow$ 
1 emission. The most important observations are obtained with the James Clerk 
Maxwell Telescope\footnote{The James Clerk Maxwell Telescope is operated by 
the Joint Astronomy Centre, Hilo (Hawaii), on behalf of the UK Particle 
Physics and Astronomy Research Council, the Netherlands Organization for 
Scientific Research (NWO) and the National Research Council of Canada.} (JCMT) 
and ISO, and are presented in the following sections after the NRAO results.

\subsection{NRAO Observations}

Single dish observations of the HCO$^+$ 1-0 and CS 2-1 lines were made in June
1996 with the NRAO 12m telescope on Kitt peak
\footnote{The National Radio Astronomical Observatory is operated by
Associated Universities, Inc., under contract with the U.S.\ National
Science Foundation.}. The adopted receiver was the
3mm SIS dual channel mixer. For the back end, the Hybrid Spectrometer was
placed in dual channel mode with a resolution of 47.9~KHz (0.16~km s$^{-1}$).
The beam width is $63''$~FWHM and the main beam efficiency
$\eta_{\rm MB}=0.86$. Pointing is accurate to $10''$ in azimuth and $5''$ in
elevation. A 
weak detection of HCO$^+$ 1 $\rightarrow$ 0 was found, but the CS 2 
$\rightarrow$ 1 line was not detected (see Table 2).

\subsection{JCMT Observations}

Observations of the $^{12}$CO 2 $\rightarrow$ 1 and 3 $\rightarrow$ 2, 
$^{13}$CO 2 $\rightarrow$ 1 and 3 $\rightarrow$ 2, C$^{18}$O 2 $\rightarrow$ 
1, HCO$^+$ 3 $\rightarrow$ 2, and [C I] (492 GHz/609 $\mu$m)
line transitions were
performed in three different wavelength bands.
The front-end receivers, the telescope beam sizes, and the efficiencies at the
various frequencies are shown in Table 1.
The main beam efficiencies $\eta_{\rm MB}$ were determined from observations
of planets by the JCMT staff.
All observations were obtained in the 125 MHz 
configuration, corresponding to a channel width of 0.078 MHz. For these high 
resolution spectra, the Digital Autocorrelating Spectrometer (DAS) at the 
back-end of the receiver was used. All observations were made in position 
switching mode, using an offset of 30$'$, 45$'$ or 90$'$ in azimuth. In this 
mode, the telescope integrates at the source position for 30 seconds and then 
switches to an emission-free position on the sky. The difference between these 
two signals yields the actual source contribution, assuming the atmospheric 
conditions do not vary significantly on the time scale of a minute. Low
order polynomial baseline fits were adopted in the reduction.

\begin{center}
\begin{tabular}{l c c c}
\multicolumn{4}{c}{Table 1}\\
\multicolumn{4}{c}{Technical Details of the JCMT}\\
\tableline
\tableline
Receiver & Frequency & $\eta_{\rm MB}$ & $\theta_{\rm beam}$ ($''$)\\
\tableline
RxA2 &230 GHz &0.69 $\pm$ 0.03 &21 $\pm$ 2\\
RxB3i &345 GHz &0.58 $\pm$ 0.02 &14.1 $\pm$ 0.5\\
RxC2 &460 GHz &0.52 $\pm$ 0.05 &10.9 $\pm$ 0.5 \\
\tableline
\tableline
\end{tabular}
\end{center}

Table 2 shows an overview of the line transitions observed at the offset
position (0$''$,0$''$), and Figure 1 shows the $^{12}$CO 2 $\rightarrow$ 1 and
3 $\rightarrow$ 2 line emission maps. Most line profiles can be fitted with a 
Gaussian at $V_{\rm LSR} = -4.9$ km s$^{-1}$, although in several spectra
a smaller peak at $V_{\rm LSR} = -5.9$ km s$^{-1}$ also occurs.
This study will 
focus on the emission at $V_{\rm LSR} = -4.9$ km s$^{-1}$. For Ced 201, the 
$^{12}$CO line profiles do not exhibit the characteristic double-peaked shape 
of self-absorption, except for a few offset positions, located near the edge
of the mapped regions, which are not associated with the central PDR.
Therefore, it can be concluded that the illuminating star is located between
the observer and the central cloud, heating the gas from the outside.

\begin{center} 
\begin{tabular}{l c c c c c}
\multicolumn{6}{c}{Table 2}\\
\multicolumn{6}{c}{Submillimeter Observations of Ced 201}\\
\tableline
\tableline
Line &I &T$_{peak}$$^{\rm a,b}$ &$\Delta$V & V$_{\rm LSR}$ &r.m.s.$^{\rm c}$\\
&K km s$^{-1}$ &K &km s$^{-1}$ &km s$^{-1}$ &K\\
\tableline
HCO$^+$ 1 $\rightarrow$ 0 & 1 & 0.15 & 0.85 & --5 & 0.03\\
CS 2 $\rightarrow$ 1 & $\ldots$ & $< 0.03$ & $\ldots$ & $\ldots$ & 0.03\\
\tableline
$^{12}$CO 2 $\rightarrow$ 1     &21.3   &23.2   &0.86   &--4.9  &0.53\\
                &7.6    &6.8    &1.1    &--5.9  &0.53\\
$^{12}$CO 3 $\rightarrow$ 2     &25.3   &33.1   &0.72   &--4.9  &0.94\\
                &7.8    &6.8    &1.1    &--5.9  &0.94\\
$^{13}$CO 2 $\rightarrow$ 1     &5.4    &6.9    &0.73   &--4.9  &0.71\\
                &1.7    &1.5    &1.1    &--5.9  &0.71\\
$^{13}$CO 3 $\rightarrow$ 2     &3.9    &5.5    &0.66   &--4.9  &1.0\\
                &$\ldots$   &$<$0.64 &$\ldots$  &--5.9  &0.64\\
C$^{18}$O 2 $\rightarrow$ 1     &$\ldots$ &$<$0.09 &$\ldots$ &$\ldots$ &0.09\\
HCO$^+$ 3 $\rightarrow$ 2       &$\ldots$ &$<$0.08 &$\ldots$ &$\ldots$ &0.08\\
\ [C I] (609 $\mu$m) &$\ldots$ &$<$1.3 &$\ldots$ &$\ldots$  &1.3\\
\tableline
\tableline
\multicolumn{6}{l}{$^{\rm b}$ In case of no detection, the upper limit,
obtained by Hanning smoothing}\\ 
\multicolumn{6}{l}{once, is shown.}\\
\multicolumn{6}{l}{$^{\rm c}$ Measured per channel of 0.078 MHz.}\\
\end{tabular}
\end{center}

\subsection{ISO Observations}

Observations of the fine-structure lines of neutral oxygen [O I] and 
ionized carbon [C II], as well as the pure rotational transitions of molecular 
hydrogen have been performed by ISO. The LWS (Clegg et al.\ 1996) AOT02
grating mode was used to obtain
spectra of the [O I] (63 $\mu$m), [O I] (145 $\mu$m) and [C II] (158 $\mu$m)
fine structure lines, whereas the H$_2$ S(1) and S(3) 
transitions were obtained using the SWS (de Graauw et al.\ 1996) AOT02
grating mode. The [C II] and H$_2$ 
observations show unresolved line profiles, while the [O I] spectra only yield 
upper limits, as summarized in Table 3. The reduction was performed with
the LWS version 6.0 pipeline software and the SWS reduction package
(Thi 1997, private communication). The
detection of the S(1) and S(3) pure rotation H$_2$ lines indicates warm
($T>200$ K) gas. This result will be further explored through the
physical models constructed below.

\begin{center}
\begin{tabular}{c c c}
\multicolumn{3}{c}{Table 3}\\
\multicolumn{3}{c}{ISO Observations of Ced 201}\\
\tableline
\tableline
Line & I & $\theta_{\rm beam}$ \\
& W m$^{-2}$ & \\
\tableline
\ [C II] (158 $\mu$m) & $3.80 \times 10^{-15}$ & 90$''$\\
\ [O I] (63 $\mu$m) & $<8\times 10^{-15}$ & 90$''$\\
H$_2$ S(1) (17 $\mu$m) & $1.10 \times 10^{-16}$ & 14$''$ $\times$ 27$''$ \\
H$_2$ S(3) (9 $\mu$m) & $5.61 \times 10^{-17}$ & 14$''$ $\times$ 20$''$ \\
\tableline
\tableline
\end{tabular}
\end{center}

\subsection{Comparison with Other Lines of Sight}

Before discussing the construction of specific models for Ced 201, it is
good to compare the observed [C II] 158 $\mu$m line intensity, which accounts
for 50\% of the total cooling rate in moderate density PDRs
($n_{\rm H}<10^4$ cm$^{-3}$),
with other lines of sight. The above value translates into $2\times 10^{-5}$
erg s$^{-1}$ cm$^{-2}$ sr$^{-1}$, which should be compared to
$2.2\times 10^{-3}$ for W3 (Boreiko, Betz, \& Zmuidzinas 1993)
and $4.6\times 10^{-4}$ for S140 (Minchin et al.\ 1994). Because the [C II]
flux scales roughly linear with the impinging UV flux in this regime, care
should be taken in comparing
these numbers. It will be shown below that the strength of the incident
radiation in interstellar units is $G_0=200$ for Ced 201, roughly equal to
the S140 value, whereas W3 is characterized by $G_0\sim 5\times 10^3$.
With these numbers one finds that the Ced 201 [C II] line is underluminous.
The remaining difference between the various lines of sight, and the topic of
this work, is the shape of illuminating radiation field.

Since the effective temperature of the illuminating star is 10,000 K, whereas
it is $\sim 30,000$ K for S140 and W3, one would
expect Ced 201 to be well described by cool PDR models. These exhibit a
much smaller [C II]/CO ratio for the higher rotational levels of CO, $J\ge 3$
(Spaans et al.\ 1994). The reasons are that 1) the soft radiation field causes
lower temperatures and hence smaller collisional excitation rates, and 2)
the abundance ratio C$^+$/CO is much larger in a soft UV field since
C ionization and CO dissociation occur in the 912-1110 \AA\ wavelength range.
Indeed, the [C II] line flux is not only small in Ced 201, which by itself
may be due to abundance, temperature, and (column) density effects, it is also
small compared to the observed CO 3-2 line with a ratio of about 100. This
is roughly an order of magnitude smaller than the corresponding ratios for the
other lines of sight. The fact that the high temperature [O I] 63 $\mu$m line,
at an excitation temperature of about 228 K above ground,
is not detected at all toward Ced 201, is consistent with
these results if the gas temperature is smaller than 150 K throughout most of
the cloud (see the PDR models below).

\section{A Physical Model for Ced 201}

The analysis of the data and construction of a physical model for Ced 201 will
proceed in several steps. First,
the average density, temperature and abundances of observed
species are determined from the observations through a simple excitation
model. The chemical abundances are then reproduced through iteration of more
elaborate 1D and 3D PDR models, which incorporate the chemical and thermal
balance of the medium, and the geometry of the reflection nebula. The
motivation for the latter is that the spatial variations in the temperature
and chemical abundances strongly influence the observed line intensities.

The intensities of emission lines are determined by the level populations
and optical depth along the line of sight.
A one-dimensional escape probability radiative 
transfer code has been applied to determine these quantities for the
detected lines. The code includes collisional and radiative
(de-)excitation processes, and computes
the level populations and line optical depth for a 
given set of physical parameters: density, temperature, column density,
line width, and the temperature of the cosmic microwave background.
Comparison with the observed intensities 
then results in a first guess for the beam-averaged kinetic 
temperature, density and column density of the observed species. This 
radiative transfer code and the procedure for analyzing the data has been
described in Jansen, van Dishoeck \& Black (1994) and Jansen (1995).
In Table 4 the derived
physical parameters are summarized. The kinetic temperature and 
the density of the main collision partner, molecular hydrogen, have been
determined using the line intensities of the $^{12}$CO and $^{13}$CO 
transitions. The optically thick $^{12}$CO lines were used to constrain the
kinetic temperature, i.e.\ this is the temperature of the (self-shielding)
molecular gas, whereas the $^{13}$CO 3-2/2-1 line ratio was used as the
main density probe (see Figure 8a of Hogerheijde et al.\ 1995).

\begin{center}
\begin{tabular}{l c}
\multicolumn{2}{c}{Table 4}\\
\multicolumn{2}{c}{Physical Parameters and Column Densities of Ced 201}\\
\tableline
\tableline
$T$                   & $(40 \pm 10)$ K\\
$n({\rm H}_2)$        & $(5000 \pm 1000)$ cm$^{-3}$\\
\tableline
$N({\rm CO})$         & $(2 \pm 1) \times 10^{17}$ cm$^{-2}$\\
$N(^{13}{\rm CO})$    & $(3 \pm 1) \times 10^{15}$ cm$^{-2}$\\
$N({\rm CI})$          & $< 2 \times 10^{16}$ cm$^{-2}$\\
$N({\rm CII})$        & $(4 \pm 2) \times 10^{17}$ cm$^{-2}$\\
$N({\rm C}^{18}{\rm O})$    & $< 4 \times 10^{13}$ cm$^{-2}$\\
$N({\rm HCO}^+)$      & $(8 \pm 3) \times 10^{11}$ cm$^{-2}$\\
$N({\rm CS})$         & $< 1 \times 10^{12}$ cm$^{-2}$\\
$N({\rm OI})$          & $<4\times 10^{18}$ cm$^{-2}$\\
\tableline
\tableline
\end{tabular}
\end{center}

It is important to realize that the C$^+$ column density is quite sensitive
to the temperature structure of the cloud due to its excitation temperature of
92 K. For the 1D and 3D models we shall therefore compare to the observed
flux directly. Since the more detailed models discussed below provide a good
fit to the observed line strengths, and indicate column densities quite similar
(within the errors) to the one above, we will use this knowledge and the value
of the C$^+$ column density to derive now the gas phase carbon abundance.

The gas phase carbon budget
is dominated by $^{12}$CO, C and C$^+$. 
The sum of the derived column densities yields for the column density of gas 
phase carbon $N(^{12}{\rm C}) = ( 6 \pm 2) \times 10^{17}$ cm$^{-2}$. If it is
assumed that the central gas cloud of Ced 201 is spherical, then one can
compare the gas phase carbon column density with the column density of 
hydrogen nuclei ($N_{\rm H}$) along the line of sight. For spherical
symmetry, $N_{\rm H} = n_{\rm H} \times diameter = (7.6 \pm 2.3) \times
10^{21}$ cm$^{-2}$, where the diameter from the $^{12}$CO 2-1 map is $120''$
at 420 pc, and the error is dominated by the non-circularity of the map.
Thus, one finds for the gas phase carbon abundance 
$A({\rm C}) = (1.2 \pm 0.4) \times 10^{-4}$. The cosmic abundance of carbon is 
$\approx 4 \times 10^{-4}$. Note here that the cosmic abundances are thought
to be 30\% lower than the Solar abundances (c.f.\ Meyer 1997).
One finds for Ced 201 that the fraction of carbon in the gas phase is
35\% $\pm$ 10\%. For
diffuse clouds a gas phase carbon abundance of $1.4\times 10^{-4}$ is derived
based on GHRS/HST observations (Cardelli et al.\ 1993), consistent with the
result found here.

If one assumes a carbon
abundance of $10^{-4}$, that C$^+$ is the dominant carbon bearing species
in the atomic PDR zone, and that the ambient density is above the critical
density of $\sim 3\times 10^3$ cm$^{-3}$ (LTE) as motivated by the detected
CO emission, then the size of the emitting
region and the energy rate per hydrogen nucleus required to reproduce the
observed line flux can be estimated. Clearly, various combinations of size
and energy rate per H atom will yield identical line luminosities. Still, in
LTE one cannot increase the emitted flux by a large amount unless the ambient
temperature is below the excitation temperature, in which case the line would
be quite weak. Therefore, taking a size of
$\sim 120''$ from the CO and IRAS dust maps, one cannot make the [C II]
emitting region much smaller than this without violating the observations,
i.e.\ above 92 K per collision one increases mostly the ambient temperature
and not the line luminosity. From these
considerations it follows (to first order) that
the beam filling factor of the emitting gas is roughly unity
and the ambient temperature in the [C II] emitting region is not less than the
excitation temperature of 92 K.

\subsection{The 1D PDR Model}

Previous studies of PDRs 
illuminated by a (interstellar) radiation field with an effective temperature
of about 30,000 K have been performed by 
Tielens \& Hollenbach (1985); van Dishoeck \& Black (1988);
Burton et al.\ (1990); Hollenbach et al.\ (1991); le Bourlot et al.\ (1992);
St\"orzer, Stutzki, \& Sternberg 1996;
at various levels of sophistication. Spaans et al.~(1994) extended PDR
calculations to softer ($T_{\rm eff}=20,000-6,000$ K) radiation fields in order
to study the changes in the photo-electric heating efficiency.
In this work two different models valid
for clouds illuminated by an intense but soft radiation field, the 1D
PDR model and the 3D PDR model, will be applied to, and tested for, Ced 201. 

The one-dimensional PDR model is an extension of the diffuse cloud 
model of van Dishoeck \& Black (1986) and is thoroughly described in Jansen et 
al.~(1995, also referred to as the Leiden PDR model). In this model, the
PDR is described as an infinite plane 
parallel slab in thermal and chemical equilibrium with a radiation field
incident from one side. The calculation of the 
chemical conditions in the PDR is performed in spatial steps, where the 
radiation field at a certain point is determined by calculating the 
attenuation by dust grains, and optical depth effects in the UV absorption 
lines of H$_2$ and CO which lead to dissociation. Moreover, shielding of CO 
by overlapping H$_2$ absorption lines (mutual shielding) is also taken into 
account (van Dishoeck \& Black 1988). The chemical network
includes 215 different species, based on 24 elements and the isotopes of H, C 
and O (Jansen et al.\ 1995).
The thermal balance is calculated simultaneously with the chemistry. The 
heating processes include photo-electric heating by dust grains and PAHs 
(Bakes \& Tielens 1994), heating by cosmic rays ($\zeta = 5.0 \times
10^{-17}$ s$^{-1}$), H$_2$ formation heating, carbon photo-ionization
heating, and collisional de-excitation of vibrationally excited
H$_2$ (Tielens \& Hollenbach 1985;
Sternberg \& Dalgarno 1989). Cooling is provided by 
spontaneous decay of collisionally excited fine-structure levels of C$^+$, C 
and O, and rotational levels of CO.

\subsection{The 3D PDR Model}

In addition to the one-dimensional PDR model, the three-dimensional 
model developed by Spaans (1996), hereafter referred to as the 3D PDR model,
is applied to Ced 201. Analogous to the 1D PDR model, this model calculates 
the thermal and chemical balance of the cool PDR, using the stellar
radiation field, 
the morphology, the elemental abundances and the (constant) density as input 
parameters. Nevertheless, differences between the models occur, due to the 
geometry, radiative transfer and chemical network.
The radiative transfer in the 3D PDR model is extended from one to three 
dimensions, thus enabling the interpretation
of line intensities from geometrically 
more complex clouds and to take into account the position of the illuminating 
star. Furthermore, the interstellar radiation field (ISRF, Draine 1978) is
explicitly taken into account as
an isotropic background radiation field illuminating the side of the cloud
opposite to the star.

In contrast with the escape probability method of the 1D PDR model, the 3D PDR
model uses Monte Carlo radiative transfer (Spaans \& van Langevelde 1992; 
Spaans 1996). According to this method, the PDR is divided in a large number 
of different cells with a typical size no larger than the mean free path of a 
photon. The physical conditions differ in each cell, thus influencing the 
excitation of atomic and molecular lines. Calculation of the radiative 
transfer proceeds by determining the stimulated emission and absorption
(line and continuum) for
propagating photon packages in each cell simultaneously. Spontaneous emission 
of line photon packages occurs in random directions.
The chemical network incorporated in the 3D PDR model is more limited, and 
includes only the $\sim$ 40 most important observable species. Nevertheless,
the network is shown to be detailed enough to model the C$^+$/C/CO transition 
accurately (Spaans \& van Dishoeck 1997).

\section{Results}

Since the central gas cloud and the illuminating star are located in
the same direction, the abundances as a function of visual extinction, 
calculated by the models, can be directly converted into column densities by 
integrating over the total extent along the line of sight. The aim of the
constructed models is to 
reproduce the observed column densities, in particular the gas phase carbon 
column density $N(^{12}{\rm C})$, the [C II] and [O I] line intensities,
and to determine the best fit input parameters of the 1D and 3D models.

Table 5 summarizes the input parameters resulting in the best fit 1D PDR 
model. The illuminating radiation field is determined by considering the 
central star as a black body with $T_{eff} =$ 10,000 K, placed at a distance
of $R_{nebula} = 0.4$ pc. This corresponds to a value of $G_0=200$ in units
of the average interstellar radiation between 13.6 and 2 eV (Spaans et al.\
1994), which is derived from the spectral type of the star and the infrared
observations presented by Casey (1991) in a straightforward manner.
The density of hydrogen nuclei in Ced 201 is found to
be $n_{\rm H} = 2n({\rm H}_2) + n({\rm H}) = 1.2 \times 10^4$ cm$^{-3}$ and
reproduces well the observed chemical abundances. The best fit column density
of molecular hydrogen $N({\rm H}_2) = 1.6 \times 10^{21}$ cm$^{-2}$ is a
measure of the physical size of the cloud. When compared with the spherically
symmetric case, $N({\rm H}_2)_{\rm sph.symm.} = 3.8 \times 10^{21}$ cm$^{-2}$,
its value indicates
that the PDR is flattened along the line of sight. Note that this also implies 
that the abundance of gas phase carbon must be higher than
35\%. In fact, the flattened geometry yields 80\% for the abundance of
carbon in the gas phase relative to the cosmic abundance.
This revised gas phase abundance is thus a factor of two higher than
the value of $1.4\times 10^{-4}$ found for diffuse clouds.
The best fit model also constrains the gas phase abundances $\delta$
of several other
elements, which are shown in Table 5. The gas phase fraction of nitrogen
is rather arbitrarily chosen, since no observations of nitrogen bearing
species were obtained.

\begin{center}
\begin{tabular}{l c}
\multicolumn{2}{c}{Table 5}\\
\multicolumn{2}{c}{Input Parameters for the Best}\\
\multicolumn{2}{c}{Fit of the 1D PDR Model}\\
\tableline
\tableline
$R_{\rm nebula}$ & 0.4 pc\\
\tableline
$n_{\rm H}$  & $1.2 \times 10^{4}$ cm$^{-3}$\\
$N({\rm H}_2)$ & $1.6 \times 10^{21}$ cm$^{-2}$\\
\tableline
$\delta$C   & 0.80\\
$\delta$O   & 0.80\\
$\delta$S   & 0.01\\
$\delta$N   & 0.70\\
$\delta$metals & 0.02\\
\tableline
\tableline
\end{tabular}
\end{center}

The 3D PDR model is applied consistently with the 1D PDR model. That is, the
only input parameter of the 3D PDR model which can be modified for a constant
density cloud is the three-dimensional geometry of the cool PDR. The best
reproductions of the observed column densities are obtained for an oblate
ellipsoid with an axis ratio of 2.4.
The 1D and 3D models which yield the best fit to the observations are
presented in Table 6.

\begin{center}
\begin{tabular}{l c c c}
\multicolumn{4}{c}{Table 6}\\
\multicolumn{4}{c}{Comparison Between the Observed Column Densities,}\\
\multicolumn{4}{c}{the 1D PDR Model and the 3D PDR Model}\\
\tableline
\tableline
Species & Observed & 1D PDR & 3D PDR Model \\
        & cm$^{-2}$ &  cm$^{-2}$ & cm$^{-2}$ \\
\tableline
H$_2$ & - & $1.6 \times 10^{21}$ & $1.5 \times 10^{21}$\\
$^{12}$CO & $(2 \pm 1) \times 10^{17}$ & $2.3 \times 10^{17}$ & $2.1 \times 10^{17}$\\
$^{13}$CO & $(3 \pm 1) \times 10^{15}$ & $4.4 \times 10^{15}$ & $3.1 \times 10^{15}$\\
C$^{18}$O & $<4 \times 10^{13}$ & $7.2 \times 10^{13}$ & $3.1 \times 10^{13}$\\
C & $<2 \times 10^{16}$ & $5.2 \times 10^{16}$ & $3.2 \times 10^{16}$\\
C$^+$ & $(4 \pm 2) \times 10^{17}$ & $3.2 \times 10^{17}$ & $3.7 \times 10^{17}$\\
CS & $<1 \times 10^{12}$ & $6.3 \times 10^{11}$ & $5.1 \times 10^{11}$\\
HCO$^+$ & $(8 \pm 3) \times 10^{11}$ & $4.1 \times 10^{11}$ & $6.8 \times 10^{11}$\\
\tableline
\tableline
\end{tabular}
\end{center}

\subsection{Chemical Structure}

The 1D PDR model yields the abundances as functions of depth for 
approximately 200 species, of which the most important ones are plotted in 
Figure 2. The assumed isotope ratios are $[^{12}{\rm C}]/[^{13}{\rm C}] = 60$ 
and $[^{16}{\rm O}]/[^{18}{\rm O}] = 500$. At the illuminated side of Ced 201,
ionized carbon is the dominant carbon bearing species. Due to line shielding 
and attenuation by intervening dust, the ionized carbon is converted into 
neutral carbon and carbon monoxide at $A_V \sim 1$ mag (Tielens \& Hollenbach, 
1985, Black \& van Dishoeck, 1987, and references therein). Nowhere in the 
cloud does neutral carbon become the dominant carbon bearing species.

The abundances of several species determined by the 3D PDR model are shown in
Figure 3. Consistent with the 1D PDR model,
the C$^+$/C/CO transition occurs at 
an extinction of $A_V \sim 1$ mag. Due to the ISRF which illuminates the 
cloud from all directions, a second C$^+$/C/CO transition zone is present at 
$A_V \sim 2$ mag on the far side of the PDR. Atomic hydrogen is the dominant
form of hydrogen at the illuminated edge, but the shielding of H$_2$ is very
efficient and leads to a sharp transition.

\subsection{Thermal Balance and Small Grains}

Both the 3D PDR model and the 1D PDR model compute the thermal balance in Ced 
201. Figures 4 and 5 show the heating and cooling rates as functions of depth 
determined by the 1D PDR model. Except for the heating by cosmic rays, 
all heating mechanisms decrease with depth into the cloud.
Photo-electric emissions by PAHs and dust grains are the dominant heating 
mechanisms in Ced 201. The heating rates of these two processes are equal, 
although only 10 \% of the solid state carbon is incorporated into PAHs in
the best fit model. This
shows how quantum effects in PAHs can contribute significantly to the heating 
process in cool radiation fields (Bakes \& Tielens, 1994).
Cooling is provided by radiative decay of collisionally excited species. The 
cooling rates due to the main coolants are shown in Figure 5, and depend 
strongly on the abundances of these species. The observations do not provide 
good constraints on the column density of atomic oxygen, and a gas phase
fraction of $3.2\times 10^{-4}$ is therefore assumed (Meyer 1997).

The thermal balance is established at an equilibrium temperature profile
determined by the heating and cooling rates. Figures 6 and 7 show these
profiles for the 1D PDR model and the 3D PDR model, respectively. The 
temperature profiles calculated by both models are approximately the 
same. The temperature at the illuminated edge is 168 K according to the 3D PDR
model, and 169 K for the 1D PDR model. The temperature decreases with 
increasing depth, and is for both models $\sim 40$ K at $A_V = 1$ mag, where 
the temperature tracer carbon monoxide becomes abundant. According to the 
1D PDR model, the temperature gradually declines at the dark 
side of the PDR. Due to the ISRF however, the 3D PDR model predicts that the 
temperature will increase again, after reaching its minimum of 22 K at $A_V = 
1.8$ mag, and becomes 34 K at the dark side of Ced 201.

For the best fit models to the chemical abundances, one can determine
the model dependent, i.e.\ with the distance to the star fixed by the
infrared observations, contribution to the heating rate from PAHs and VSGs.
The models indicate that 10\% of the gas phase carbon is locked up in the
form of PAHs, i.e.\ $D_{\rm PAH}=0.1$, where PAHs are defined as any
carbonaceous particle of linear size less than 15\AA.
Additional models were run without any contribution of PAHs and VSGs to check
the robustness of this result. The resulting thermal balance in the latter
case yields temperatures at the edge of the cloud which are lower by 50\%, and
strongly underestimates the observed [C II] flux.
The fraction $D_{\rm PAH}=0.1$ also yields good agreement with the
observed CO excitation
temperatures. The precise value for $D_{\rm PAH}$ depends on the model.
Nevertheless, a firm conclusion is that heating by PAHs and VSGs is required
to explain the observed properties of Ced 201, provided no heating sources
other than photo-electric heating and cosmic rays are important.

A final crucial comparison comes from the predicted [C II] 158 $\mu$m and
[O I] 63 $\mu$m line
intensities, because they reflect the thermal balance of the medium most
strongly. In the 1D case one finds values of $1.1\times 10^{-5}$ for [C II]
and $1.9\times 10^{-5}$ erg s$^{-1}$ cm$^{-2}$ sr$^{-1}$ for [O I]. The 3D
results are systematically higher and yield $1.8\times 10^{-5}$ and
$2.7\times 10^{-5}$ erg s$^{-1}$ cm$^{-2}$ sr$^{-1}$, respectively.
Comparison with the observed [CII] intensity would
thus favor the 3D model, but the difference is small.

\subsection{Pure Rotational H$_2$ Lines}

The H$_2$ J=3 and J=5 levels in the vibrational ground state lie at
$\sim 1,000$ K and $\sim 2,500$ K above ground,
and therefore trace warm molecular gas.
The ISO observations of the H$_2$ S(1) and S(3) pure rotational transitions 
are not included in the PDR models, since at kinetic temperatures of $T < 100$ 
K, i.e.\ in a large part of Ced 201, the absolute intensities of these lines
are negligible. For higher temperatures these lines are useful temperature 
tracers because they are optically thin and the level 
populations are in thermal equilibrium. The ratio between the observed 
transitions is $I(9 \mu {\rm m})/I(17 \mu {\rm m}) = 0.69$, which indicates
a kinetic temperature of $T \sim 330$ K (Black \& van Dishoeck 1987).

This observational result is clearly at variance with the model
predictions. Particularly since the effective temperature of the
illuminating is low. It appears that an additional heating source is required
in some part of the cloud. If we assume that a 
thin layer of hot gas is present at the illuminated edge or any where else
inside the cool PDR, then it 
follows from the absolute intensities that this layer is not thicker than 
$\delta A_{\rm V} \sim 0.05$ mag for a mean H$_2$ fraction of 0.2.
We therefore feel that the value of $D_{\rm PAH}$
remains secure since the chemical constraints which enter into it, derive
from regions in the cloud with $A_{\rm V}>0.3$ mag. The upper limit of
$5\times 10^{-5}$ erg s$^{-1}$ cm$^{-1}$ sr$^{-1}$ for the [O I] 63 $\mu$m
line and its density dependence of $1.0\times 10^{-8}n_{{\rm H}_2}$
erg s$^{-1}$ cm$^{-1}$ sr$^{-1}$ for temperatures higher than 228 K and
subthermal excitation, indicate that an upper limit for
the H$_2$ gas density of $5\times 10^3$ cm$^{-3}$ when a
unity beam filling factor is adopted.
>From the CO maps we estimate a beam dilution factor of not more than two,
so the upper limit is consistent with the CO density estimate and vibrational
de-excitation heating by H$_2$ can be ruled out.

Furthermore, the continuum intensity observed by IRAS is around
$2\times 10^{-3}$ erg s$^{-1}$ cm$^{-2}$ sr$^{-1}$, whereas the [C II]
intensity is $2\times 10^{-5}$ erg s$^{-1}$ cm$^{-2}$ sr$^{-1}$. This
translates into a heating efficiency for the gas of 1\%, a generic value for
PDRs. The temperature of 300 K itself requires a $G_0$ of $5\times 10^3$
for an ambient H$_2$ density of $5\times 10^3$ cm$^{-3}$, incompatable with the
value $G_0=200$ from the illuminating star. All in all, the high
temperature molecular gas remains to be explained, and does not seem to be
accommodated by the process of photo-electric heating. In fact, observations
presented by Witt et al.\ (1987) on the scattering properties of dust in
Ced 201 indicate that there is a narrow size distribution of grains skewed
toward larger than average particle sizes. This renders a larger contribution
to the photo-electric heating by small grains unlikely.

One might argue that the H$_2$ levels are populated through UV pumping.
Even though the reflection nebula is relatively close to the illuminating
star, the low effective stellar temperature strongly quenches the 912-1110 \AA\
flux relative for H$_2$ fluorescence. A straightforward calculation shows that
UV irradiation can account for at most 20\% of the S(3) line, thereby rendering
it a minor contribution.

An additional heating source like turbulent dissipation could be present
in the cool PDR (Falgarone \& Puget 1995). This requires the input of
kinetic energy on the scale of the cloud, possibly through a weak
(3-5 km s$^{-1}$) shock. A C-shock of 7 km s$^{-1}$ into a gas of pre-shock
density $10^4$ cm$^{-3}$ would also suffice, but would likely lead
to an observable [O I] 63 $\mu$m line. Finally, the case of Ced 201 appears
not to be unique. Observations obtained by Timmermann et al.\ (1996) for S140
indicate similarly warm gas. It would be quite interesting if more of such warm
regions show up in H$_2$ rotational line data.

\section{Conclusions}

The 1D PDR and the 3D PDR model reproduce the chemical abundances
and the thermal balance of the cool PDR Ced 201 as derived from the
observations. All observed lines,
obtained with the JCMT, ISO and NRAO, are accommodated by these models,
except for the pure rotational transitions of H$_2$. The observed H$_2$ S(1)
and S(3) line intensities indicate the presence of hot gas with a kinetic
temperature of $T \sim 330$ K, which is probably located in a thin
layer of $\delta A_{\rm V} \sim 0.05$ mag at the illuminated edge of the
cool PDR. The best fit models indicate that the gas phase carbon abundance
is 50\% of Solar, and that 10\% of the available carbon atoms is in the
form of PAHs and VSGs.

\acknowledgments Research in astrochemistry in Leiden is supported by the
Netherlands Organization for Scientific Research (NWO).
MS is supported by NASA through grant HF-01101.01-97A,
awarded by the Space Telescope Institute, which is
operated by the Association of Universities for Research in Astronomy,
Inc., for NASA under contract NAS 5-26555.
We are grateful to the assistence of Jante Salverda and Remo Tilanus
in obtaining the observations presented in this work. We would like to
thank Wing-Fai Thi for the ISO-SWS reduction, Byron Mattingly for
obtaining the NRAO 12m observations, and John Black for general discussions
and computer codes.

\newpage

\begin{figure}
\label{figure1}
\caption{
Contourplot of Ced 201. In the left panel the integrated line intensities with
spacings of $60''$ of the $^{12}$CO 2 $\rightarrow$ 1 transition are shown.
The contours represent values of 0.6, 5, 10, 15, 20 and 25 K km s$^{-1}$. The
right panel presents the integrated line intensities of the $^{13}$CO 2
$\rightarrow$ 1 line with spacings of $20''$. The contour values correspond
to 2, 4 and 6 K km s$^{-1}$. The $^{13}$CO 2 $\rightarrow$ 1 emission is also
indicated in white in the left panel.}
\end{figure}

\newpage

\begin{figure}
\label{figure2}
\caption{
Abundances of carbon bearing species in Ced 201
as functions of visual extinction into the cloud, for the
1D PDR model.}
\end{figure}

\newpage

\begin{figure}
\label{figure3}
\caption{
Abundances of several species as functions of
visual extinction, for the 3D model.}
\end{figure}

\newpage

\begin{figure}
\label{figure4}
\caption{
The most important heating rates as functions of visual
extinction (1D PDR model).}
\end{figure}

\newpage

\begin{figure}
\label{figure5}
\caption{
Cooling rates originating from the main coolants
as functions of visual extinction (1D PDR model).}
\end{figure}

\newpage

\begin{figure}
\label{figure6}
\caption{
Temperature distribution throughout the cloud, according
to the 1D PDR model.}
\end{figure}

\newpage

\begin{figure}
\label{figure7}
\caption{
The kinetic gas temperature as a function of
extinction, according to the 3D PDR model.}
\end{figure}

\clearpage
\centerline{\psfig{figure=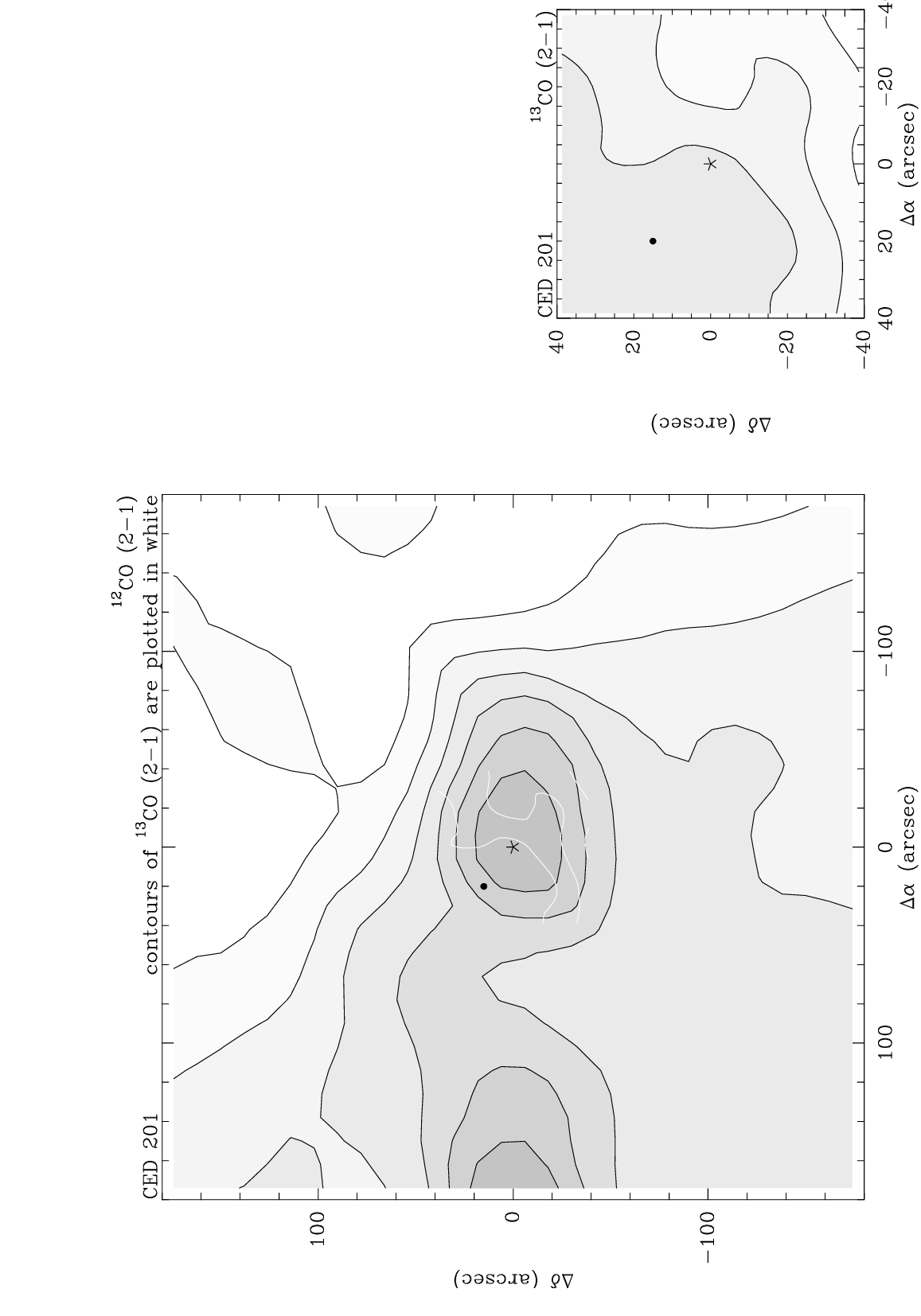,width=15.0truecm}}

\clearpage
\centerline{\psfig{figure=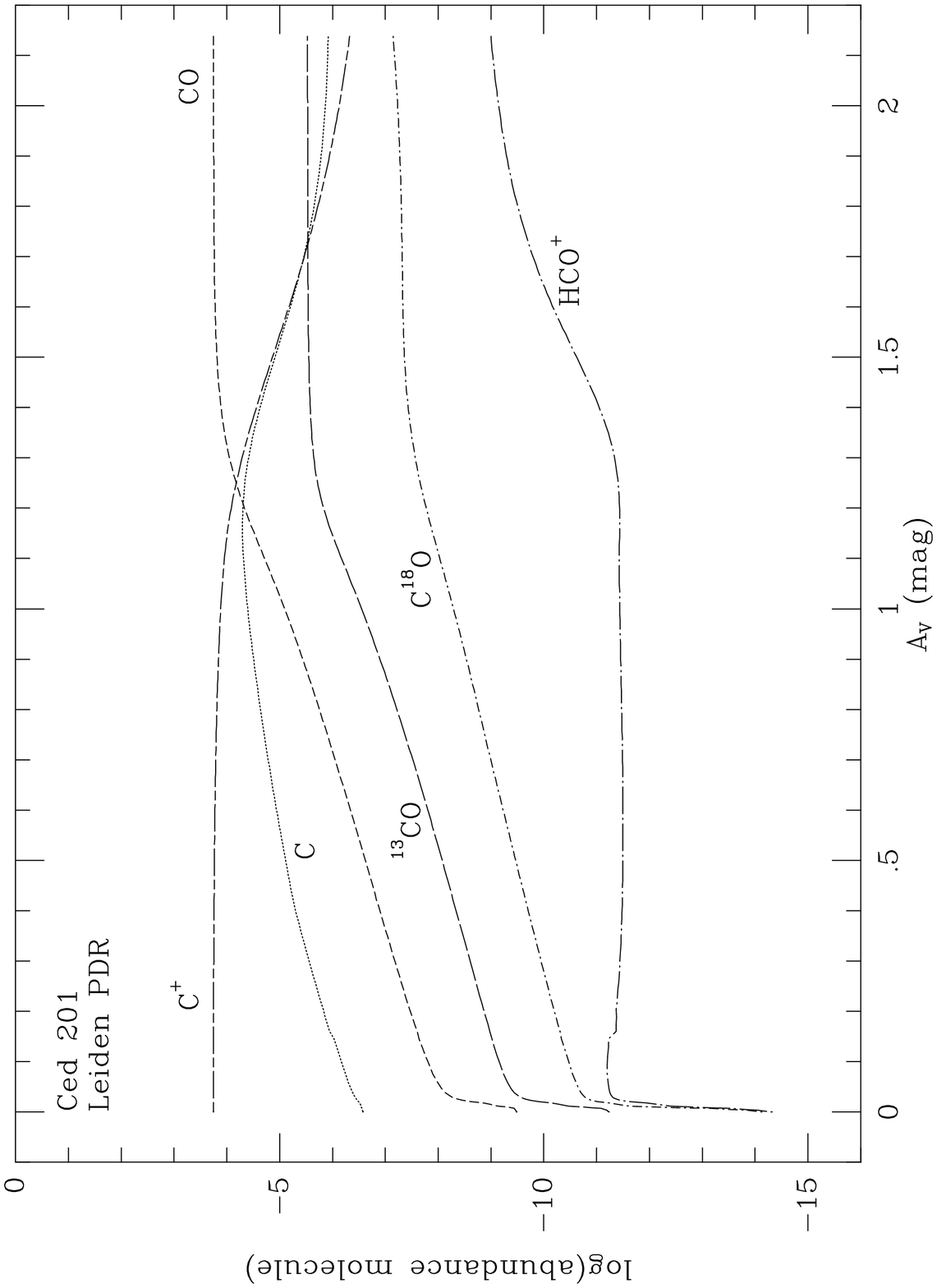,width=15.0truecm}}

\clearpage
\centerline{\psfig{figure=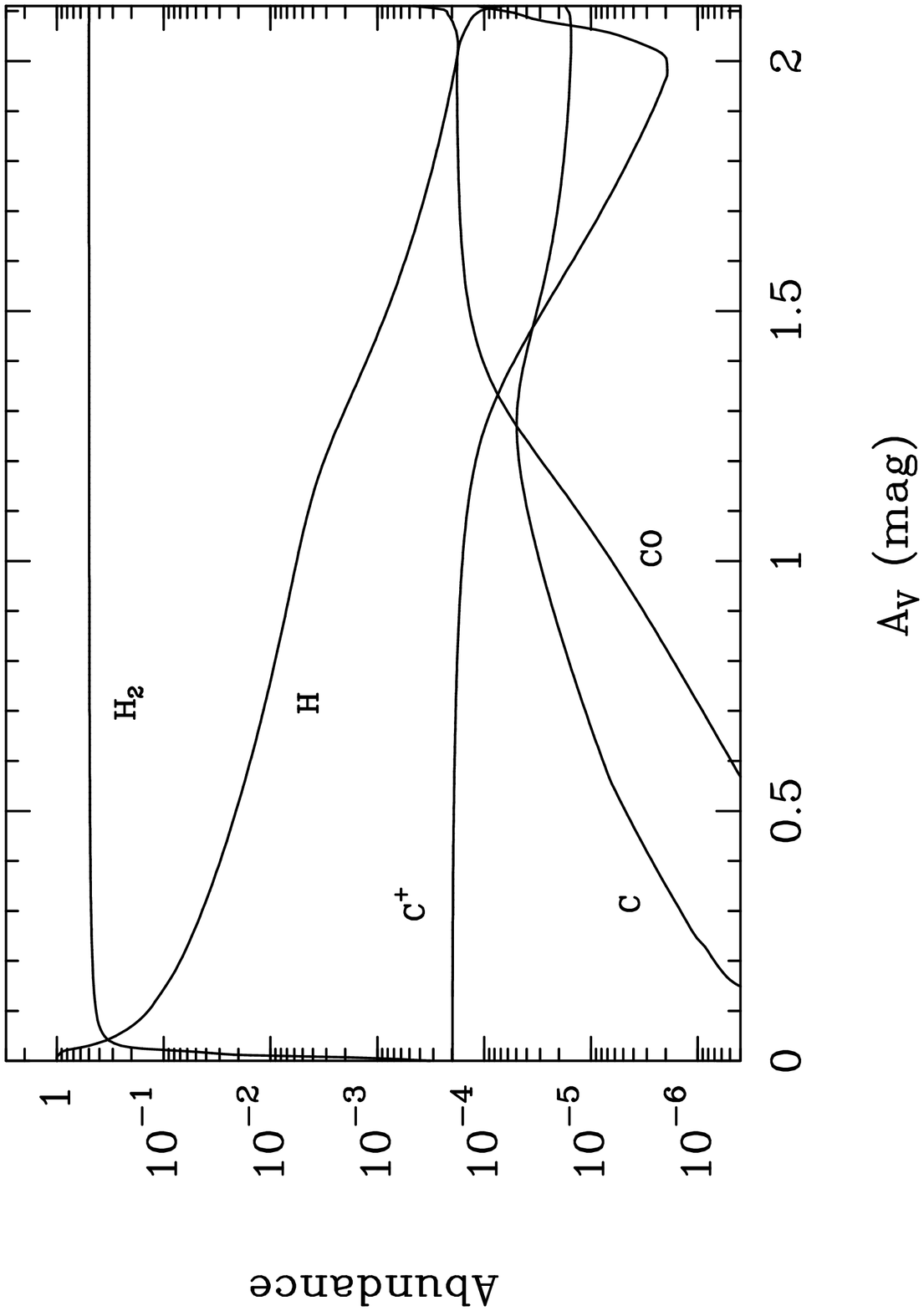,width=15.0truecm}}

\clearpage
\centerline{\psfig{figure=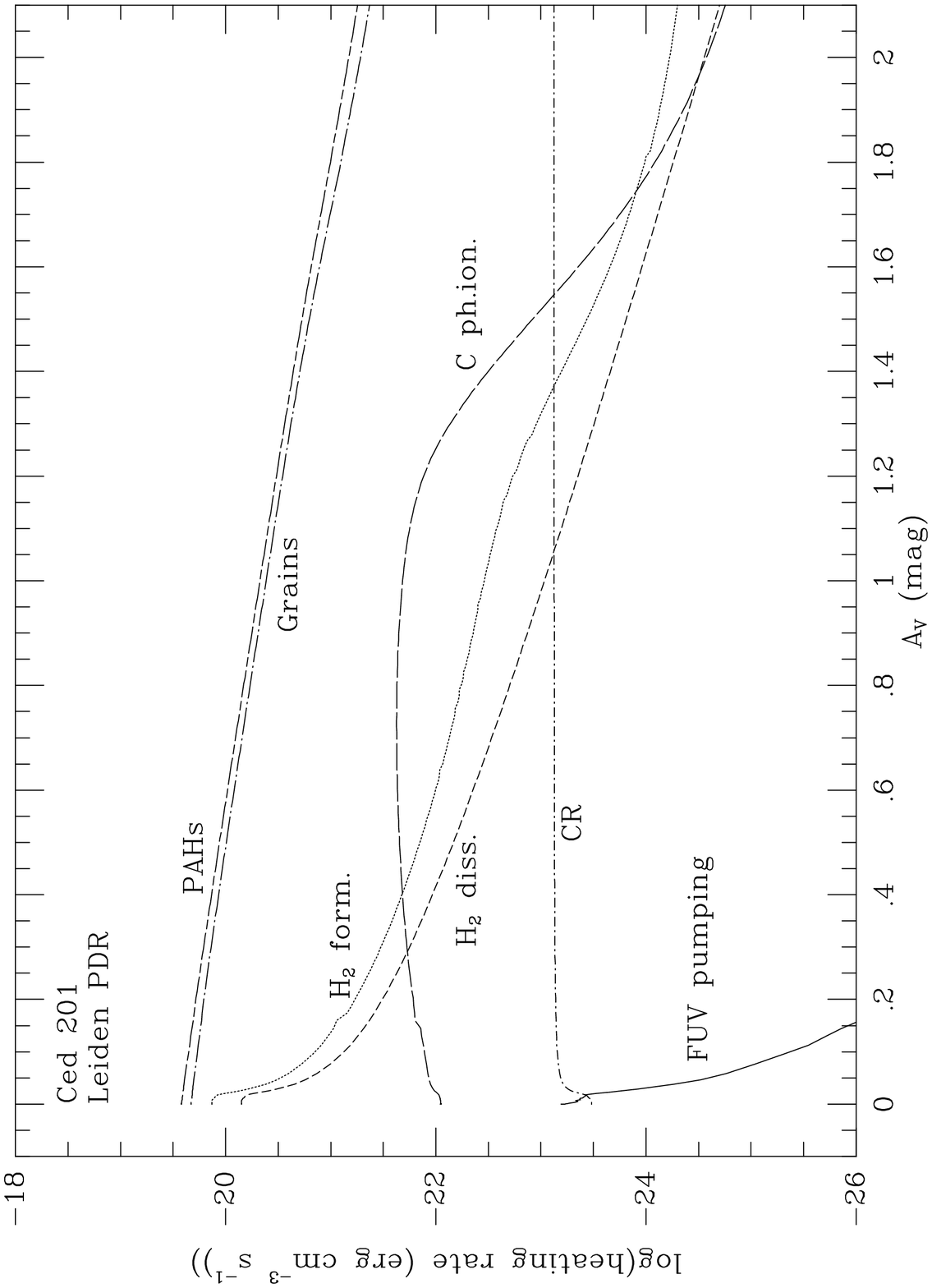,width=15.0truecm}}

\clearpage
\centerline{\psfig{figure=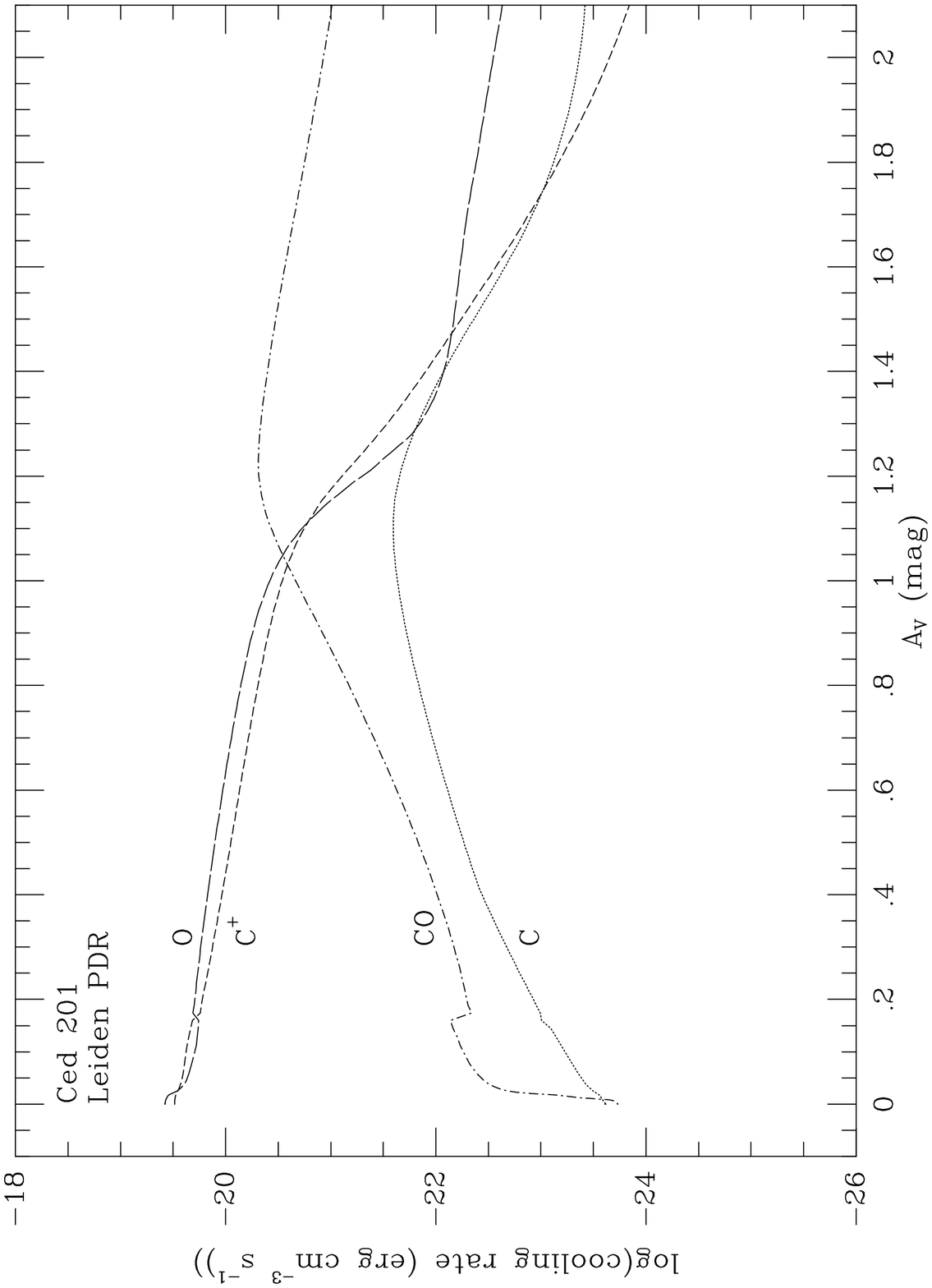,width=15.0truecm}}

\clearpage
\centerline{\psfig{figure=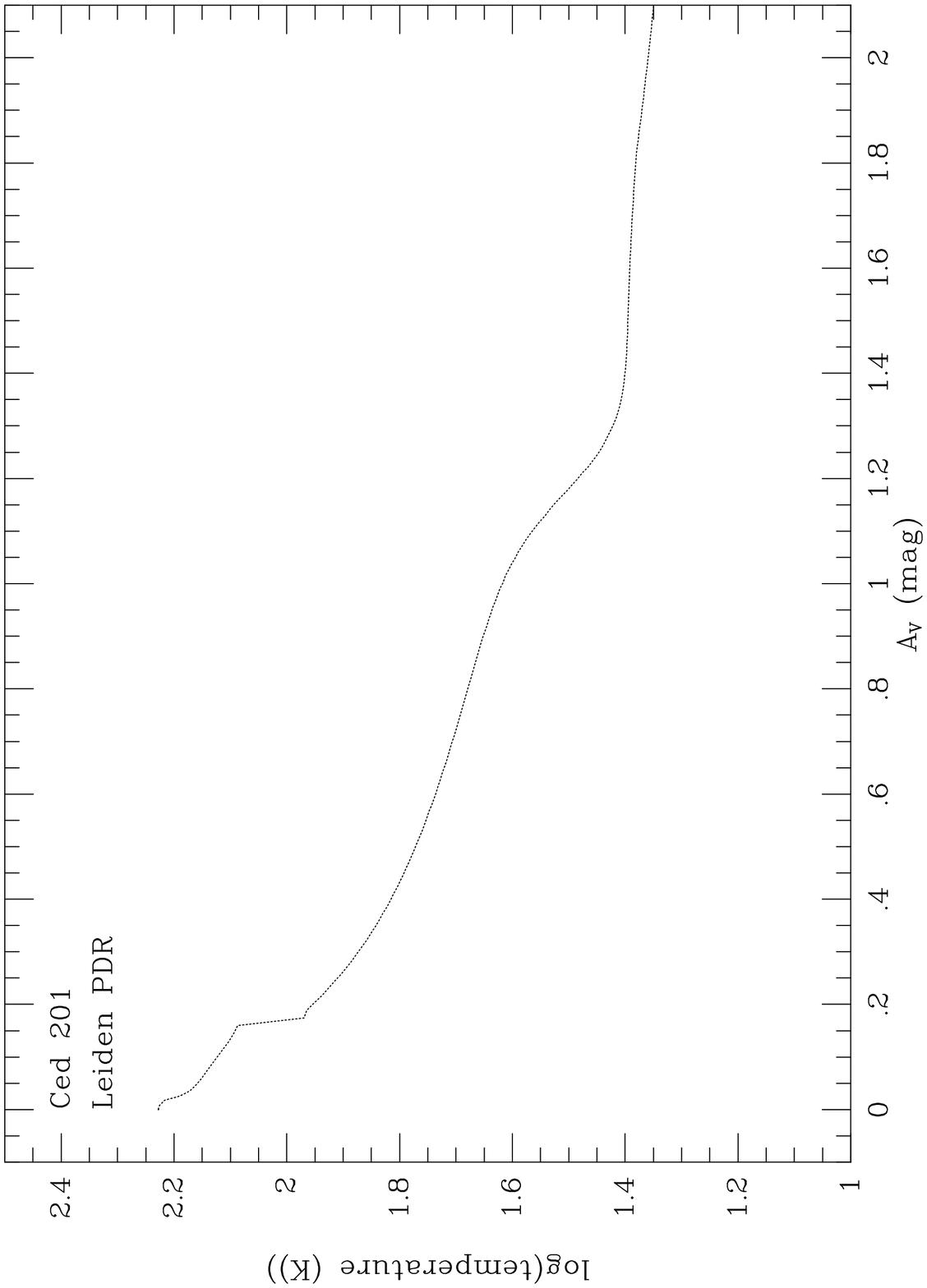,width=15.0truecm}}

\clearpage
\centerline{\psfig{figure=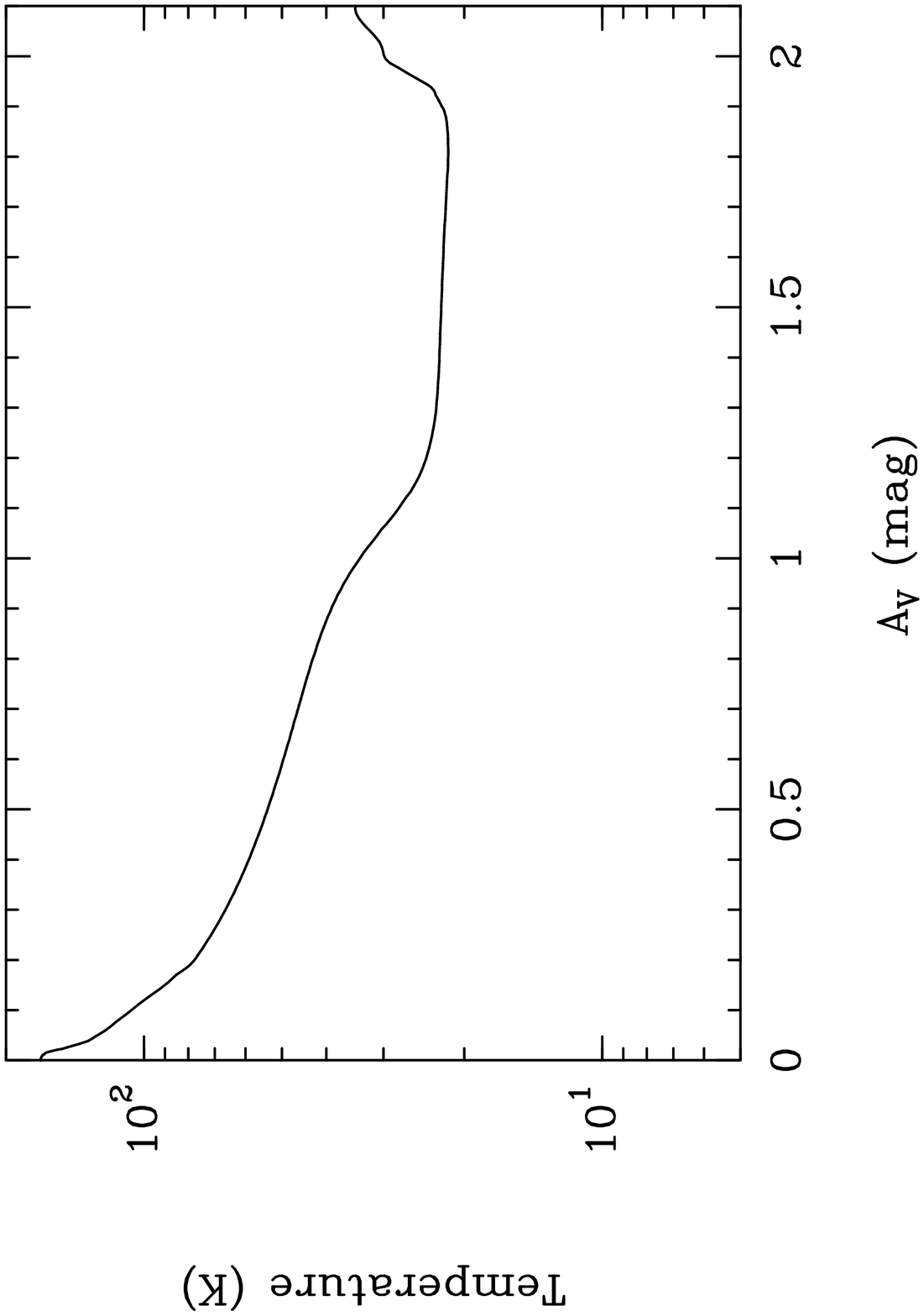,width=15.0truecm}}


\newpage
\begin{thebibliography}{}
\bibitem[]{}Bakes, E.L.O., \& Tielens, A.G.G.M.\ 1994, ApJ, 427, 822
\bibitem[]{}Boreiko, R.T., Betz, A.L., \& Zmuidzinas, J.\ 1990, ApJ, 353, 181
\bibitem[]{}Black, J.H. \& van Dishoeck, E.F.\ 1987, ApJ, 322, 412
\bibitem[]{}Burton, M.G., Hollenbach, D.J., \& Tielens, A.G.G.M.\ 1990, ApJ, 365, 620
\bibitem[]{}Cardelli, J.A., Federman, S.R., Lambert, D.L., \& Theodosiou,
C.E.\ 1993, ApJ, 416, L41
\bibitem[]{}Casey, S.C.\ 1991, ApJ, 371, 183
\bibitem[]{}Clegg et al.\ 1996, A\&A, 315, 
\bibitem[]{}de Graauw et al.\ 1996, A\&A, 315, 
\bibitem[]{}Draine, B.T.\ 1978, ApJS, 36, 595
\bibitem[]{}Falgarone, E., \& Puget, J.-L.\ 1995, A\&A, 293, 840
\bibitem[]{}Hogerheijde, M.R., Jansen, D.J., \& van Dishoeck, E.F.\ 1995, A\&A,
294, 792
\bibitem[]{}Hollenbach, D.J., \& Tielens, A.G.G.M.\ 1997,
Ann.\ Rev.\ Astron.\ Astrophys., 35, 179
\bibitem[]{}Hollenbach, D.J., Takahashi, T., \& Tielens, A.G.G.M.\ 1991, ApJ, 377, 192
\bibitem[]{}Jansen, D.J.\ 1995, Ph.D.\ Thesis, Leiden Observatory
\bibitem[]{}Jansen, D.J., van Dishoeck, E.F., Black, J.H., Spaans, M. \& Sosin, C.\ 1995, A\&A, 302, 223
\bibitem[]{}Jansen, D.J., van Dishoeck, E.F., \& Black, J.H.\ 1994, A\&A, 282, 605
\bibitem[]{}le Bourlot, J., Pineau de For\^ets, G., Roueff, E. \& Flower, D.R.\ 1992, A\&A, 267, 233
\bibitem[]{}Meyer, D.M.\ 1997, in Molecules in Astrophysics: Probes and
Processes, IAU 178, (Kluwer:Dordrecht), P.\ 407
\bibitem[]{}Minchin, N.R., White, G.J., Stutzki, J., Krause, D.\ 1994, A\&A,
291, 250
\bibitem[]{}Spaans, M.\ 1996, A\&A, 307, 271
\bibitem[]{}Spaans, M., Tielens, A.G.G.M., van Dishoeck, E.F. \& Bakes, E.L.O.\ 1994, ApJ, 437, 270
\bibitem[]{}Spaans, M. \& van Dishoeck, E.F.\ 1997, A\&A, 323, 953
\bibitem[]{}Spaans, M. \& van Langevelde, H.J.\ 1992, MNRAS, 258, 159
\bibitem[]{}Sternberg, A., \& Dalgarno, A.\ 1989, ApJ, 338, 197
\bibitem[]{}St\"orzer, H., Stutzki, J., \& Sternberg, A.\ 1996, A\&A, 310, 592
\bibitem[]{}Tielens, A.G.G.M., \& Hollenbach, D.J.\ 1985, ApJ, 291, 722
\bibitem[]{}Timmermann, R., Bertoldi, F., Wright, C.M., Drapatz, S., Draine,
B.T., Haser, L., \& Sternberg, A.\ 1996, A\&A, 315, L281
\bibitem[]{}van Dishoeck, E.F., \& Black, J.H.\ 1986, ApJS, 62, 109
\bibitem[]{}van Dishoeck, E.F., \& Black, J.H.\ 1988, ApJ, 334, 771
\bibitem[]{}Witt, A.V., Graff, S.M., Bohlin, R.C., \& Stecher, T.P.\ 1987, ApJ,
321, 912
\end{thebibliography}
\end{document}